\documentclass[twocolumn,superscriptaddress]{revtex4-1}

\usepackage{graphicx,amsmath}

\newcommand{\be}{\begin{equation}}
\newcommand{\ee}{\end{equation}}
\newcommand{\ben}{\begin{eqnarray}}
\newcommand{\een}{\end{eqnarray}}
\newcommand{\bF}{\begin{figure}}
\newcommand{\eF}{\end{figure}}
\newcommand{\bi}{\begin{itemize}}
\newcommand{\ei}{\end{itemize}}

\date{\today}
\begin{document}
\author{George~C.~Knee}
\email{gk@physics.org}
\affiliation{Department of Physics, University of Warwick, Coventry CV4 7AL, United Kingdom}
\author{Patrick Rowe}
\affiliation{Department of Chemistry, University of Warwick, Coventry CV4 7AL, United Kingdom}
\affiliation{London Centre for Nanotechnology, Thomas Young Centre, and Department of Physics and Astronomy, University College London, 17-19 Gordon Street, London WC1H 0AH, United Kingdom}
\author{Luke~D.~Smith}
\affiliation{Department of Physics, University of Warwick, Coventry CV4 7AL, United Kingdom}
\author{Alessandro~Troisi}
\affiliation{Department of Chemistry, University of Warwick, Coventry CV4 7AL, United Kingdom}
\affiliation{Department of Chemistry, University of Liverpool, Liverpool L69 7ZD, United Kingdom}
\author{Animesh~Datta}
\email{animesh.datta@warwick.ac.uk}
\affiliation{Department of Physics, University of Warwick, Coventry CV4 7AL, United Kingdom}

\title {Structure-Dynamics Relation in Physically-Plausible Multi-Chromophore Systems}

\begin{abstract}
We study a large number of physically-plausible arrangements of chromophores, generated via a computational method involving stochastic real-space transformations of a naturally occurring `reference' structure, illustrating our methodology using the well-studied Fenna-Matthews-Olson complex (FMO). 
To explore the idea that the natural structure has been tuned for efficient energy transport we use an atomic transition charge method to calculate the excitonic couplings of each generated structure and a Lindblad master equation to study the quantum transport of an exciton from a `source' to a `drain' chromophore. We find significant correlations between structure and transport efficiency: High-performing structures tend to be more compact and, among those, the best structures display a certain orientation of the chromophores, particularly the chromophore closest to the source-to-drain vector. We conclude that, subject to reasonable, physically-motivated constraints, the FMO complex is highly attuned to the purpose of energy transport, partly by exploiting these structural motifs.
\end{abstract}
\maketitle

The process of {excitonic} energy transport is of great interest across many fields of study, from evolutionary biology to the engineering of solar cells. It is a key component of both natural photosynthesis (occurring in light harvesting complexes (LHCs), which are assemblages of chromophores found in a famously `warm and wet' environment) and artificial photovoltaics. It is tempting to suggest that, since natural selection has had perhaps billions of years~\cite{Blankenship2014} over which to improve the efficiency of the process, we might learn from nature how to better engineer artificial light harvesting systems. This view has gathered interest in recent years, in tandem with claims that naturally occurring systems can exhibit outstandingly high efficiency~\cite{PanitchayangkoonHayesFransted2010,Scholes2010}. It is often said that quantum coherence could play a pivotal role in enhancing transport~\cite{EngelCalhounRead2007,ColliniWongWilk2010,HuelgaPlenio2013,IshizakiFleming2012} through constructive interference of excitation pathways (although this is hotly debated~\cite{WilkinsDattani2015,DuanProkhorenkoCogdell2016,BaghbanzadehKassal2016a}) and that the structural arrangement of chromophores has been {optimised for this functionality}~\cite{Scholes2010}.  {Contrariwise, it} might be suggested that efficient transport through a network of chromophores is actually generic and most of the possible structural arrangements would show similar performance. It is difficult to draw conclusions about the general structure-dynamics relationship from studying only a handful of {naturally occurring LHCs}.  Indeed, while it is clearly desirable to learn from the results of natural selection~\cite{Schlau-Cohen2015}, we must also be able to relax the constraints that it operated under in order to uncover the potential for technological advantage, while at all times keeping our imagination on a short enough leash so that any new structures remain physically plausible. 

Theoretical efforts have so far proceeded along two paths. The first technique is to randomly `sprinkle' point-like electrical dipoles into a restricted volume: This approach has allowed the use of genetic algorithms to optimise their positions~\cite{ScholakWellensBuchleitner2011a}, and later showed how fast transport is typically aided by a `backbone plus pair' geometry by reinforcing constructive interference~\cite{MostardaLeviPrada-Gracia2013}, and correlated to the centrosymmetry of the Hamiltonian describing the energy of the system~\cite{ZechMuletWellens2014}.  The spatial density of dipoles was considered~\cite{JesenkoZnidaric2012}, and (although generally correlated with efficient exciton transport) was found to saturate to the densities typically found in natural LHCs~\cite{MohseniShabaniLloyd2013}. A drawback of such approaches is that they may not have much to say about realistic systems, whose structure and properties are determined by more complicated, highly-constrained relations between anisotropic molecules extended in physical space. 
 
 {The second technique begins with a} commonly accepted mathematical model of an existing system, and performs perturbations on it.  {This approach has enabled investigations of robustness and identification of dominant transport pathways in a naturally occurring LHC~\cite{BakerHabershon2015},  as well as transport efficiency in} purple bacteria compared with counterfactual structures, obtained through rotating ~\cite{BaghbanzadehKassal2016a} or `trimming'~\cite{BaghbanzadehKassal2016} chromophores from the LH1 and LH2 complexes. {The latter works speculate on the} relative importance of coherent supertransfer in these systems (noting that it is a `spandrel' -- or evolutionary byproduct -- rather than an adaption) as well as the degree of attunedness of the original structure for energy transport.  A possible drawback of these approaches, as we explain below, is {the potential physical implausibility of the perturbed models}, especially when the {transformations involve `deleting' coupling terms or making} uncorrelated perturbations to the Hamiltonian matrix.
Other strategies {deploy abstract models of the transport system~\cite{MohseniRebentrostLloyd2008,CarusoChinDatta2009,SchijvenKohlbergerBlumen2012,Caruso2014,LiCarusoGauger2015} to learn general principles} from idealised `network' models that may operate to a greater or lesser extent in realistic systems. 

In this Letter, {we incorporate the best of the above strategies to address two key questions on the structure-dynamics relationship for exciton transport: (i) which structural motifs influence exciton dynamics in typical physically-plausible multi-chromophore complexes? and (ii) do naturally occurring complexes appear to be attuned for certain transport tasks (by exploiting these motifs or otherwise)? {We proceed by simulating the transport properties of a large number of artificial, or \emph{ersatz} structures} obtained from a reference one by geometrical perturbations (GPs). 

\begin{figure}[]
\includegraphics[width=8.25cm]{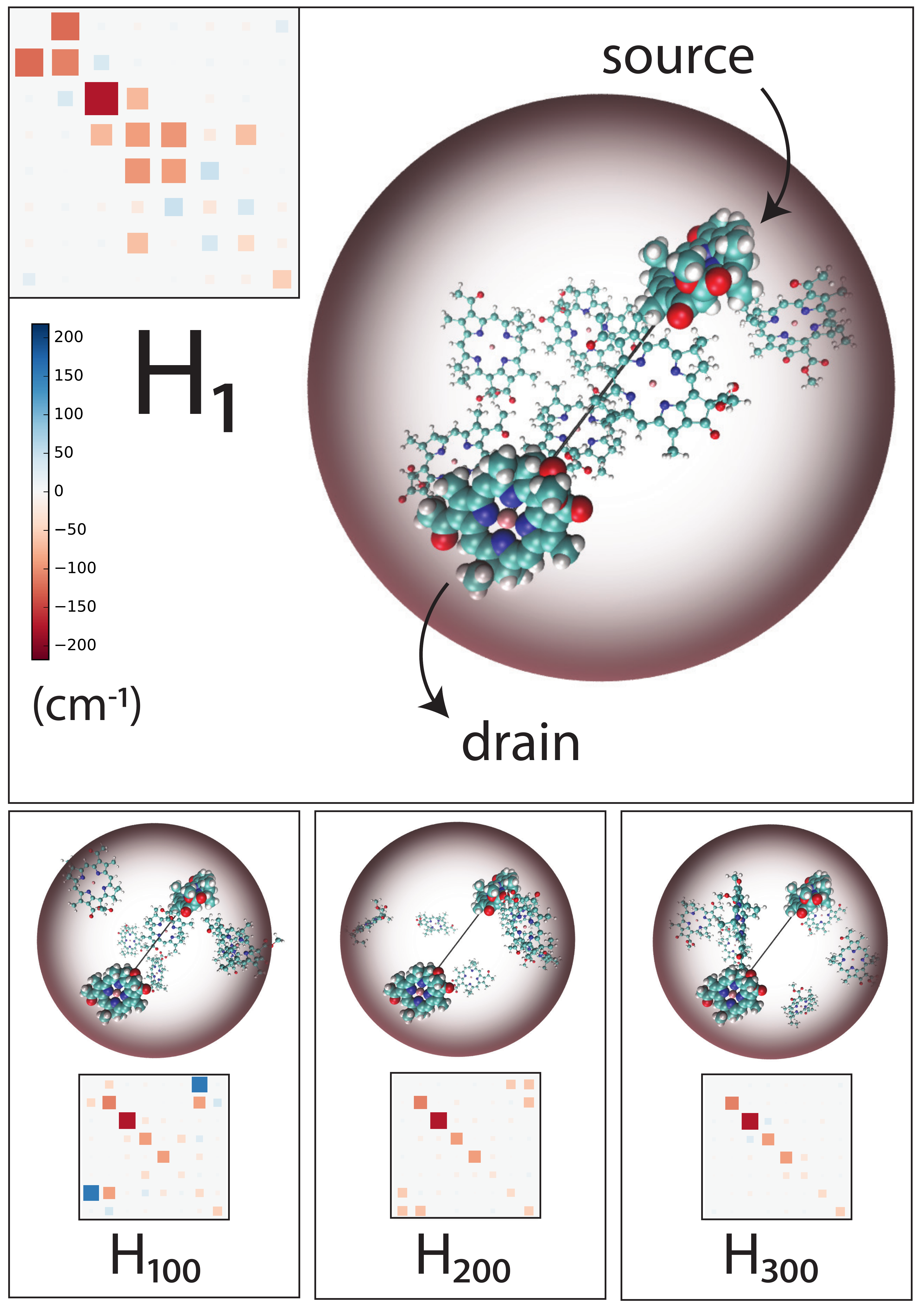}
\caption{\label{structures}The FMO trimer is composed of three (monomer) subunits, each consisting of eight BChl chromophores supported in a protein scaffolding. We consider a single unit of the complex, where the protein is retained only implicitly through its role as a dielectric and vibrational environment. Observe our enclosing sphere, which all chromophores must fit within, and the `transport vector' which connects {BChls} 1 and 3 (enlarged for clarity), which serve as source and drain respectively. We generate 50,000 new structures through physical changes to the FMO. Three representative structures are shown, along with their resulting Hamiltonian matrices (in the chromophore basis) which we calculate using an Atomic Transition Charge method. To aid clarity, the energies of all BChls have been shifted such that the energy of BChl 1 is zero. \label{fig:main}}
\end{figure}


{We generate} a sample of plausible structures, associating to each an $N\times N$ Hamiltonian matrix representing a network of $N$ chromophores where a single excitation can exist. {We take the naturally occurring Fenna-Matthews-Olson (FMO) complex~\cite{FennaMatthews1975} as our reference structure, chosen for its simplicity and relevance as a model system for LHCs~\cite{LohnerAshrafCogdell2016} (although our methodology could however just as well be applied to any LHC). Present in species of green sulfur bacteria, the complex takes the form of three identical protein monomers, each of which comprising a protein matrix which contains {$N=8$} identical bacteriochlorophyll-a (BChl) chromophores~\cite{MoixWuHuo2011} (see Fig. 1). {As we will be interested in exciton transport between BChls 1 (`source', closest to the antenna) and 3 (`drain', closest to the reaction centre), we fix their position, while the other 6 BChls are rigidly displaced and rotated by a sequence of $n=50,000$ Monte-Carlo moves, starting with their reference locations~\cite{TronrudWenGay2009,3EOJ,BermanWestbrookFeng2000}. During the GP, we retain only plausible \emph{ersatz} structures by rejecting those where} the minimum inter-atomic separation between the BChls is smaller than $d_{\text{min}}=2\mbox{\normalfont\AA},$ a cut-off intermediate between bonding and van der Waals distances; and contain the geometric centroid of each BChl within a sphere of radius $r_{\text{max}} = 22.3\mbox{\normalfont\AA}$ from the midpoint between BChls 1 and 3 (roughly representing the volume contained by the protein matrix in the reference structure).  

To each \emph{ersatz} {structure $i$} we associate the excitonic Hamiltonian
\begin{equation}
{H_i} = \sum_{l=1}^{{N=8}} \epsilon_l|l\rangle\langle l | + \sum_{m\neq n} V^{(i)}_{mn}\Big(|m\rangle\langle n | +|n\rangle\langle m | \Big).
\end{equation}
This expression is written with respect to an orthonormal set of basis states $|l\rangle$, each representing the situation where BChl $l$ is excited and all other BChls are not.  The first structure $i=1$ is our reference structure. {The diagonals $\epsilon_l$  represent the energy landscape for non-interacting chromophores. While some previous studies set their distribution to be uniform~\cite{ScholakWellensBuchleitner2011a}, we retain a non trivial distribution by fixing the diagonals to those corresponding to the reference FMO (see supplementary information S4 and S6).
An energy gradient, aided by a bath-mediated relaxation, can `funnel' excitations towards lower-lying excited states.  Although our modelling of the bath forbids this relaxation, we argue that these `site' energies can be nevertheless very influential on the exciton dynamics. They determine the energy eigenstates of $H_i$ and therefore influence the phase relationship and subsequent interference between different pathways through the network. }.}

The off-diagonal elements $V^{(i)}_{mn}$ arise from Coulomb interactions between transition charge densities, and depend on the mutual orientation of the chromophores.  It is well known that a simple point-dipole approximation to $V_{mn}^{(i)}$ would break down at the short distances we consider~~\cite{CurutchetMennucci2017}: we therefore adopt a superior approach ~\cite{MadjetAbdurahmanRenger2006}, partitioning the atomic transition density for each structure $i$ into atomic transition charges $q^{t}_{\alpha m}$ (with coordinates $\mathbf{R}_{\alpha m}^{(i)}$) centred on atom $\alpha$ of chromophore $m$. The Coulombic interaction is then summed to evaluate the coupling  
\begin{equation}
V_{mn}^{(i)}=f\sum_{\alpha}^{N_{a}}\sum_{\beta}^{N_{b}}\frac{q^{t}_{\alpha m}q^{t}_{\beta n}}{|\mathbf{R}_{\alpha m}^{(i)}-\mathbf{R}_{\beta n}^{(i)}|}.
\label{couplings}
\end{equation}
The matrix elements $V_{mn}^{(i)}$ in Eqn. (2) are scaled by the factor $f=0.8$ to capture the dielectric effect of protein and solvent~~\cite{AdolphsMuhMadjet2007}.
 
{We use a phenomenological dynamical model for the motion of a single exciton through our network of chromophores. While several variations~\cite{WuLiuMa2012} and more elaborate models exist~\cite{Chenu2015}, our Lindblad master equation captures the essential effects of dephasing and dissipation, and represents a completely positive transformation~\cite{BreuerPetruccione2007} on the density matrix $\rho$. 
Defining $\mathcal{L}(\rho) \equiv L\rho L^\dagger-\frac{1}{2}\{L^\dagger L,\rho\}$, we have
\begin{align}
\dot{\rho_i} = &-i[H_i,\rho_i] +\nonumber\\
& \mathcal{L}^{\textrm{sink}}(\rho_i) +\sum_{l=1}^8 \left[\mathcal{L}^{\textrm{deph}}_l(\rho_i)+\mathcal{L}^{\textrm{diss}}_l(\rho_i)\right].
\label{me}
\end{align}
The first term represents closed-system dynamics, and the other terms represent effects arising from interaction with an environment, with the collapse operators taken to be independent of the structure index $i$ and given by 
\begin{align}
L^{\textrm{sink}} &= \sqrt{2\Gamma^\text{sink}}\quad|\text{sink}\rangle\langle 3|;\quad \!\!\!\!\!\!&\Gamma^\text{sink} &= 6.3 \textrm{ps}^{-1}\nonumber\\
L^{\textrm{deph}}_l &= \sqrt{2\Gamma^\text{deph}}\quad|l\rangle\langle l|;\quad &\Gamma^\text{deph}& = 2.1 \textrm{ps}^{-1}\nonumber\\
L^{\textrm{diss}}_l &= \sqrt{2\Gamma^\text{diss}}\quad|\text{env}\rangle\langle l|;\quad &\Gamma^\text{diss} &= 0.0005 \textrm{ps}^{-1}.
\label{collapse}
\end{align}
We have expanded the Hilbert space to include $|\text{sink}\rangle$ and $|\text{env}\rangle$ (environment) which are states accessible through dissipation but which cannot build up any coherence with the rest of the network. {Here, }$\mathcal{L}^{\textrm{sink}}$ is a fast and irreversible process that represents successful capture of the exciton at BChl 3, where it is transferred to a reaction centre, or `sink'~\cite{Blankenship2014}. All other incoherent processes operate independently of the chromophore index $l$. {$\mathcal{L}^{\textrm{diss}}$ represents the relatively slow process of exciton dissipation (or decay).} The $\mathcal{L}^{\textrm{deph}}_l$ term represents the {averaged coupling of BChl $l$ to its vibrational environment, leading to a local loss of coherence.} If this happens too quickly, the {exciton} can be `frozen' and prevented from {evolving as per the Zeno} effect~\cite{MisraSudarshan1977}. If dephasing {is too slow},  destructive interference can lock the {exciton} in a particular subspace~\cite{CarusoChinDatta2009,WuSilbeyCao2013}. {We choose rates extracted from spectroscopy experiments at 77K~\cite{RebentrostMohseniAspuru-Guzik2009}, but }our conclusions are robust to moderate changes in these values (see supplementary information S1). 

Although the choice of a realistic initial condition for the exciton transport is the subject of some debate~\cite{KassalYuen-ZhouRahimi-Keshari2013}, {we set $\rho_i(0)=|1\rangle\langle 1 |,$ an exciton localised on {BChl} 1 (our `source'). This is consistent with the accepted photo-excitation process, which occurs in the nearby chlorosome (which acts as an antenna~\cite{Blankenship2014}).  We numerically solve the Lindblad master equation for $\rho_i(t)$, which can be thought of as a Hilbert-space operator `trajectory' associated with each \emph{ersatz} structure. It is this association which will allow us to explore the structure-dynamics relationship.}

\begin{figure}[t]
\centering
\includegraphics[width=8.25cm]{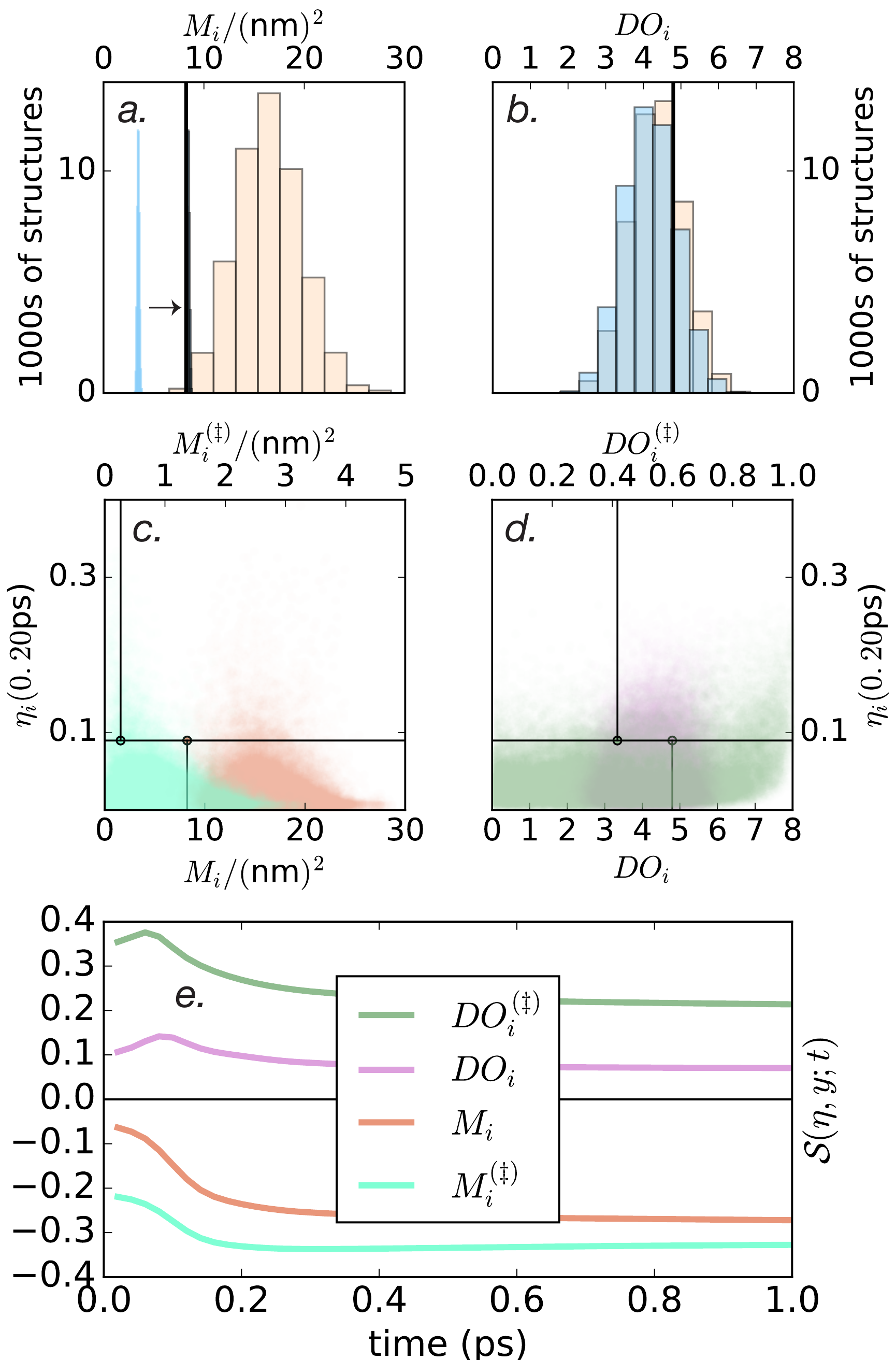}
\caption{The upper panels show histograms for a) moment of inertia ($M_i$) and b) dipole orientation ($DO_i$) of the geometric perturbation (GP, peach) and rotations only (GP(R), blue) sample of structures. Reference values $M_1$ and $DO_1$ are shown with thick black lines. In panel a) the very narrow histogram for GP(R) has been shifted by 5 (nm)$^2$ for clarity. The middle panels show scatter plots of our structural properties versus transport efficiency at $0.2$ ps. One can visualise c) a negative correlation with moment of inertia $M_i$ (orange, bottom axis), and a stronger negative correlation with the key moment of inertia $M^{(\ddag)}_i$ (turqoise, top axis); and d) a positive correlation with $DO_i$ (purple, bottom axis) along with a stronger positive correlation with the key dipole orientation $DO_i^{(\ddag)}$ (green, top axis). In panel d), we have frozen the moment of inertia of all structures to be that of the reference $M_i\rightarrow M_1$. In each panel, the data point for the reference structure is shown with a black crosshair. These time-resolved correlations are shown in panel e). Note the separation of timescales: up to about $1/2\Gamma^{\text{deph}}\sim 250$fs, the correlations change rapidly before stabilising. \label{figQ1}}
\end{figure}

First we consider how compact the chromophores are along the `transport vector' which connects the source and drain (see Fig.~(\ref{fig:main})). Define 
\be
M_i \equiv \sum_{k\in\textrm{Mg}} (r^{(i)}_k)^2,
\ee
(the mass-normalised moment of inertia), where $r^{(i)}_k$ is the perpendicular distance of each atom in {structure} $i$ to the transport vector and the sum is taken only over the coordinates of the Magnesium atoms that lie roughly in the centre of each BChl. Fig.~(\ref{figQ1}a) shows that the vast majority of our sample has $M_i>M_1$. We infer that it is difficult to pack the BChls into a volume smaller than nature has, without having unreasonably short distances between chromophores. We will return to this point later on.

Our focus is on investigating the structural properties of those randomly generated structures which perform `best' at some exciton transport function, e.g. fast or high-yield transport; high-retention storage or even fast energy-dissipation. Those structures performing well can be inspected for characteristics which might form the basis of design principles for synthetic applications. Here, we quantify the performance of our dynamics via the transport efficiency
\be
 \eta_i(t) \equiv \langle \textrm{sink}|\rho_i(t)|\textrm{sink}\rangle,
\ee
which measures both the speed and yield of energy transfer. 

To identify the main structural motifs influencing the exciton dynamics we use the Spearman rank correlation coefficient $\mathcal{S}(x,y)$. Ranging between $-1$ and $+1$, the coefficient measures the extent to which two quantities $x$ and $y$ are monotonically related~(see supplementary information S3). In this work, we fix $x=\eta_i(t),$ and begin with $y=M_i.$ Fig.~(\ref{figQ1}c,e) shows an inverse correlation between the moment of inertia $M_i$ and the transport efficiency $ \eta_i(t)$. Structures that are more compact tend to exhibit faster transport.

In fact, this realisation has a bearing on the second of our key questions (ii): To investigate the relative performance of the reference structure, we introduce the normalised rank (henceforth `rank') $\mathcal{R}_x$ defined as
\be
\mathcal{R}_x \equiv 1-o(x)/n,
\ee
where $o(x) \in \{1,\ldots,n=50,000\}$ is the ordinal rank by measure $x$ through our sample of \emph{ersatz} structures. Thus, at time $t$, the reference structure has $\mathcal{R}_\eta$ equal to one (zero) if it has the highest (lowest) $\eta_i(t)$ among all \emph{ersatz} structures in the sample. A rank of 0.5 means there are as many \emph{ersatz} structures outperforming as underperforming the reference. The standard deviation of $\eta_i(t)$ is written as $\sigma(\eta)\equiv\sqrt{\frac{1}{n}\sum_i (\eta_i(t)-\mu)^2}$, with $\mu \equiv \sum_i \eta_i(t)/n$. It has its usual statistical definition (as the square root of the variance), and gives an idea of the overall variation in transport efficiencies exhibited by our sample of structures at each instant of time. If $\sigma(\eta)$ is small, then a high rank may correspond to a small increase in efficiency in absolute terms. Likewise a large deviation hints that there is more variation in the sample, and more efficiency at stake for high ranking structures to benefit from. 

The raw data for transport efficiency are shown in Fig~(\ref{figQ2}a), with our summary statistics shown below. Fig~(\ref{figQ2}c) shows that the standard deviation increases from zero to a maximum around 4 ps and then decreases: it also shows the high ranking of the reference structure amongst our sample of \emph{ersatzs}, which exhibits a sizeable variation of transport efficiencies. A partial explanation for this lies in the tendency for our physically-plausible GPs to generate structures with $M_i$ larger than that of the FMO, coupled with the negative correlation between $M_i$ and transport efficiency~\footnote{We made a new sample with much stricter constraints on the moment of inertia (removing the bias), and the reference FMO continued to perform well there (see supplementary information S5). }. 

\begin{figure}[b]
\centering
\includegraphics[width=8.25cm]{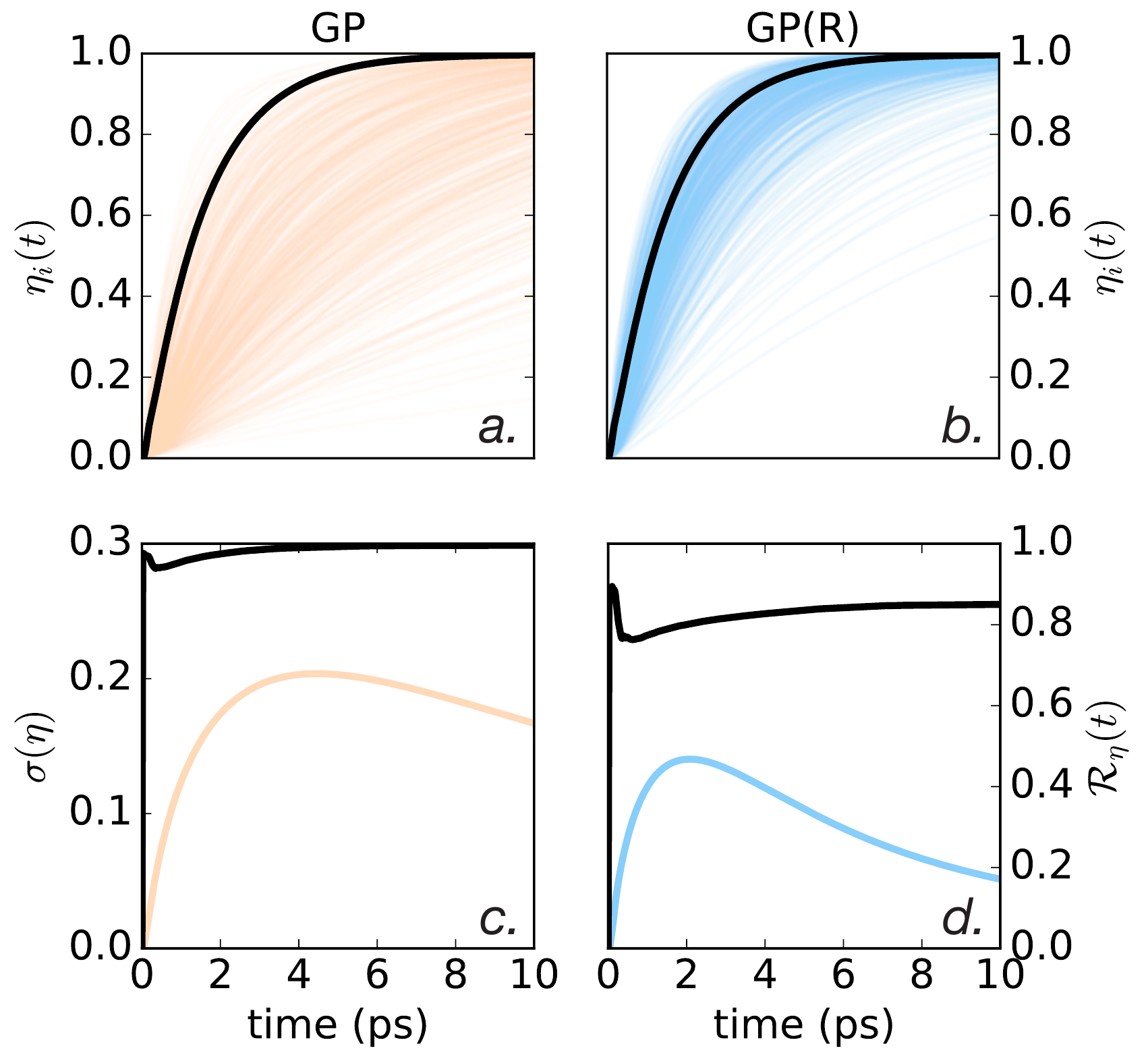}
\caption{The upper panels show exciton transport efficiency over time for a) every hundredth structure generated with geometric perturbation (GP) method (peach) and b) every hundredth structure generated with rotations only (GP(R)) method (blue). In both cases the reference FMO structure is shown with a thick black line. The lower panels show the standard deviation in efficiency $\sigma(\eta)$ of the entire sample (left axis, colour), again for c) GP and d) GP(R) methods. The rank of the reference FMO structure (right axis) is shown with a thick black line. \label{figQ2} }
\end{figure}

Delving deeper into the structure-dynamics relationship, a closer visual inspection of the top ten performing structures all feature a BChl directly between source and drain, while the bottom ten did not have this feature. This caused us to investigate the contribution of the chromophore closest to the transport vector. We call this the `key' chromophore, with index $\ddag$. In our sample each distinct chromophore assumed this role with approximately equal frequency. This step allowed us to uncover some stronger correlations with  $M^{(\ddag)}_i\equiv(r^{(i)}_\ddag)^2$ of relative importance up to $\sim3.5\times$ but generally decreasing over time, as shown in Fig.~(\ref{figQ1}e).

But does the combination of a) the strength of correlation between $M_i$ and $\eta_i$ and b) the bias in $M_i$ of our sample of \emph{ersatzs} (with respect to the reference $M_1$) completely explain the high ranking of the reference structure? We address this by studying a sample GP(R) where only rotations are allowed. For these \emph{ersatzs}, the structures are equally compact~\footnote{The moment of inertia is only approximately constant when we perform only rotations of the chromophores. This is because we rotate chromophores around their geometric centroids, but calculate moment of inertia with the central Magnesium atom. These two points are slightly displaced with respect to each other ($\sim0.312 \mbox{\normalfont\AA}$)} (Fig.~\ref{figQ1}a), but Fig.~(\ref{figQ2}b,d) shows that the rank of the reference structure has significantly decreased. This is consistent with the improved distribution of $M_i$ in the GP(R) sample, which typically leads to higher excitonic couplings (see supplementary information S7). Because the reference still ranks highly, however, we posit that there is some aspect of the relative orientation of the BChls that is also tuned up for transport. 

Investigating this further requires us to probe the structure at a finer level. We adopt an hierarchical approach by fixing variables where we have found a correlation (i.e. $M_i\rightarrow M_1$) before investigating the next variable. This way, we avoid marginalising over the correlations that are known to exist, which would otherwise wash out the evidence of further correlations. Define the total dipole orientation 
\be
DO_i = \sum_l|\mathbf{d_l}^{(i)}\cdot\mathbf{A}|,
\label{do}
\ee
as the sum of projections of the $Q_y$ transition dipole moment $\mathbf{d}_l$ of each BChl~\cite{MilderBruggemannGrondelle2010} onto the transport vector $\mathbf{A}.$ $DO_i$ is zero(one) when the dipoles are perpendicular (parallel or antiparallel) to the transport vector.  Fig.~(\ref{figQ1}e) shows a positive correlation $\mathcal{S}(\eta_i(t),DO_i)$ which again increases (by up to $\sim3.3\times$) when we select only the key chromophore in Eq.~(\ref{do}) -- i.e. $DO_i^{(\ddag)}$. 

Fig.~(\ref{figQ1}b) shows that the reference structure has a slightly larger than average $DO_i,$ which could partially explain its high rank independently of the considerations of moment of inertia. Another likely reason that the reference may be so highly ranked is our conditioning on natural FMO energies. We made a sample where the couplings $V_{mn}^{(i)}$ are fixed, but energies $\epsilon_l$ are chosen randomly, finding $\mathcal{R}(t)\approx0.5$. To summarise, given the energy distribution $\epsilon_l$ of the natural FMO, its structure is highly attuned; but given its structure, the energy distribution is not attuned. This hints at improvements that may be possible through on-site energy changes alone. The correlations found in this work are statistically reliable and significant: we demonstrate this in the supplementary information (S2 and S3) using (respectively) convergence diagnostics and a p-value analysis.

We have performed a statistical analysis of a large sample of physically-plausible mult-chromophore structures, and succeeded in extracting significant correlations between their structural properties and the time-resolved efficiency with which they transport excitons. We found that compact structures tend to exhibit higher efficiency, and among those, efficiency tended to be higher in structures whose chromophores have their transition dipole aligned with the vector pointing from source to drain. Furthermore, we provide compelling evidence that one chromophore in particular has a dominating influence on the transport efficiency -- namely the chromophore that is closest to this vector. The most influential structural property is actually the orientation of this key chromophore, although it becomes less dominant after around 100 fs. These insights could allow for improvements in the transport properties of genetically engineered excitonic networks~\cite{Park2016}. Lastly, since our reference structure -- that of the natural FMO -- is very compact and has moderately well-aligned chromophores, it has a remarkably high efficiency when compared to the other structures we generated. Although these structural properties go some way to explaining the high performance, they most likely do so in combination with energetic considerations -- namely the on-site energies of the network (which we were not able to perturb in the same physically-plausible sense as the couplings).

Our geometric perturbation approach places the natural FMO close to the top of the class, with advantages of the order of 20pp (percentage point) increase in transport efficiency around 4 ps ($\mathcal{R}_\eta(4\text{ps}) = 0.991$, or $466$th of $50,000$). This is perhaps more surprising than the striking ranking of light harvesting complexes in purple bacteria, where the 5.5 s.d. advantage may attributed to the high symmetry of the chromophore structure~\cite{BaghbanzadehKassal2016}, a property that the FMO does not ostensibly possess. Conclusions such as these (based on changes in physical space) are at odds with those inferred from other methodologies of generating \emph{ersatzs}~\cite{CarusoChinDatta2009,BakerHabershon2015}, such as uncorrelated matrix-element-perturbation (UMEP) methods. These work by perturbing the Hamiltonian directly in the Hilbert space (see supplementary information S8) and rank the natural FMO as generic with $\mathcal{R}\approx0.5$.  Given the lack of assured physical plausibility, the necessity of choosing small deviations, and the difficulty of discovering inherent biases, we view the insights gleaned from these methodologies to be of limited relevance to the question of whether natural structures are attuned.

Follow-up research may proceed along several directions: i) running a more sophisticated optimisation routine (such as those that simulate evolutionary processes~\cite{ScholakWellensBuchleitner2011a}) to find high performing structures, rather than the simple random search exhibited above; ii) more powerful, machine-powered pattern-finding could be leveraged~\cite{MostardaLeviPrada-Gracia2013} to seek structural motifs that correlate more strongly with desired transport properties than the rudimentary measures we chose; iii) applying our methodology to alternative pigment-protein complexes or solid state energy transfer materials might reveal other (distinct) structure-dynamics relationships. It should be possible to consider a variable number of chromophores, calculating (for example) the minimum number of chromophores needed to reach a certain efficiency. Lastly, it may be possible to invert the structure-dynamics link in order to estimate unknown structures from measurements of excited-state dynamics (e.g. those probed with ultrafast spectroscopy~\cite{Yuen-ZhouKrichKassal2014}), potentially complementing other theoretical techniques that explore a structure-spectrum link~\cite{ChenuCao2017}.

\acknowledgements{
G.C.K. was supported by the Royal Commission for the Exhibition of 1851, P.R. by UK EPSRC, A.T. by ERC (Grant No. 615834), and AD by UK EPSRC (EP/K04057X/2)  This work benefited from the Monash Warwick Alliance. The authors thank Rocco Fornari, Martin Plenio, Felix Pollock and Kavan Modi for interesting discussions.}
\section*{Ancillary files}
The data needed to reproduce the results of this manuscript can be found at \url{https://doi.org/10.6084/m9.figshare.c.3784874.v1}. This includes all 50,000 structures and corresponding Hamiltonians generated with the GP method.

%

\bibliography{DattaReferences,gck_full_bibliography}

\widetext
\clearpage
\renewcommand{\thesection}{S\arabic{section}}   
\renewcommand{\theequation}{S\arabic{equation}}
\renewcommand{\thefigure}{S\arabic{figure}}
\renewcommand{\bibnumfmt}[1]{[S#1]}
\renewcommand{\citenumfont}[1]{S#1}

\setcounter{equation}{0}
\setcounter{figure}{0}

\begin{center}
\textbf{\large Supplementary Information}
\end{center}

\section{Stability of results with respect to model parameters}
Although in the main text we chose what seems to be highly precise values, for example for the dephasing rate $\Gamma^{\text{deph}}=2.1$ps$^{-1}$, we show here that our conclusions would not drastically change if such parameters were changed by a small amount. We tried using $(100\%\pm \delta\%)\Gamma^{\text{deph}}$ for various values of $\delta$. The raw data are shown in Figure~\ref{stability}. Clearly a change of $\delta=-100\%$ has a significant qualitative impact on the results (since then there is no dephasing at all), but the dynamics is less affected for positive values for $\delta$. The population in $|$sink$\rangle$ changes by less than $20\%$ (relatively) when $\delta \in [-50,450]$. This covers the rate of $9.1$ps$^{-1}$ thought to represent the environment at room temperature~\cite{RebentrostMohseniKassal2009}.  
\begin{figure}[b]
\includegraphics[width=8cm]{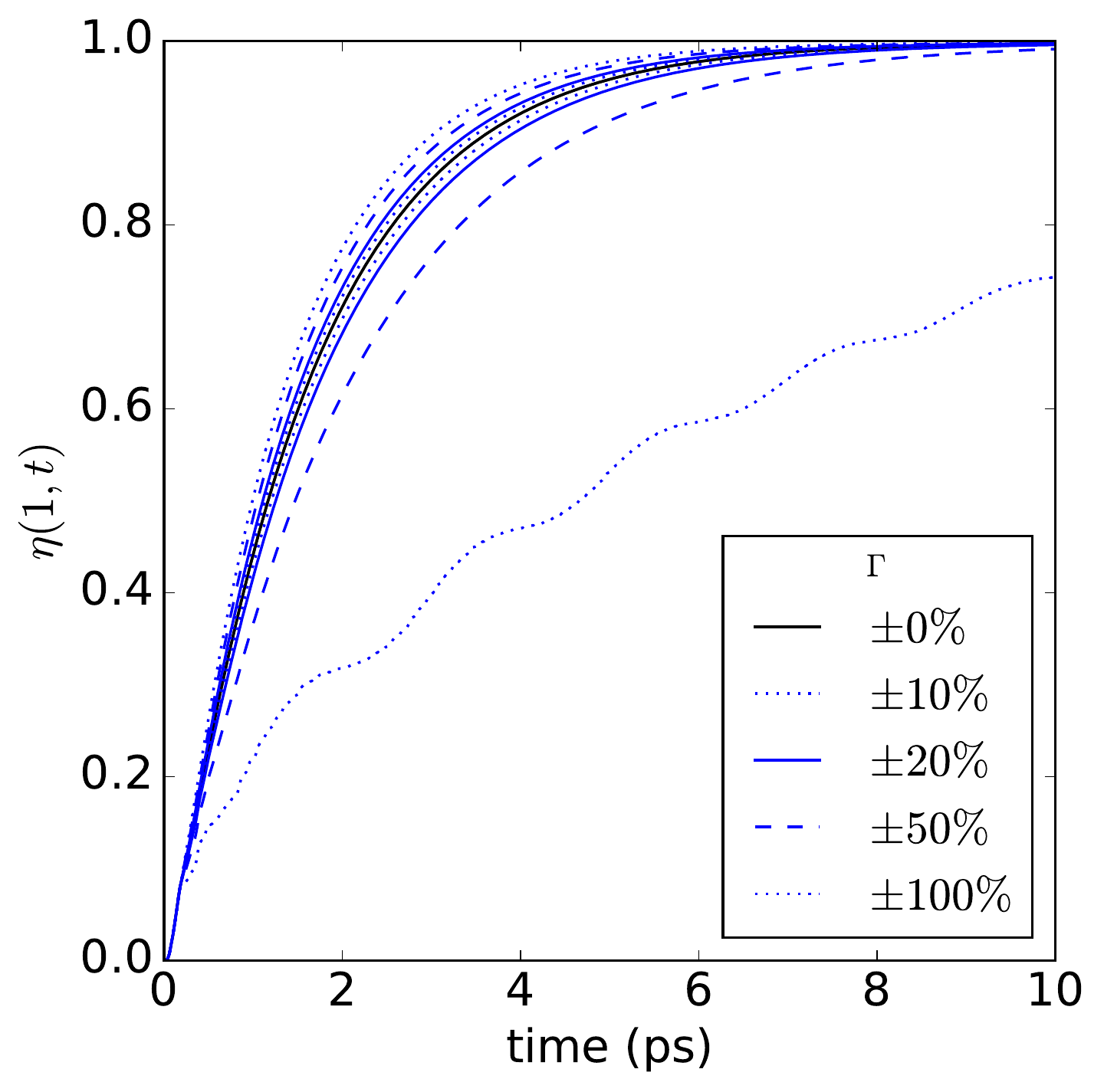}
\caption{\label{stability}Stability of the time-resolved transport efficiency (for reference geometry) with different values for the dephasing rate $\Gamma_{\textrm{deph}}$. Plot relates to the reference Hamiltonian $H_1$ (i.e. the natural FMO). }
\end{figure}

\section{Convergence diagnostics}
We would like to be able to trust that the samples we used to do our analysis are large `enough', so that the statistics we calculate will be close `enough' to those of the true distribution (approached as the sample size is taken to infinity). We plotted various quantities against library size, including the moment of inertia (Figure \ref{convergence3}) and rank and standard deviation (Figures \ref{convergence1} and \ref{convergence2}). 
\begin{figure}[t]
\includegraphics[width=8cm]{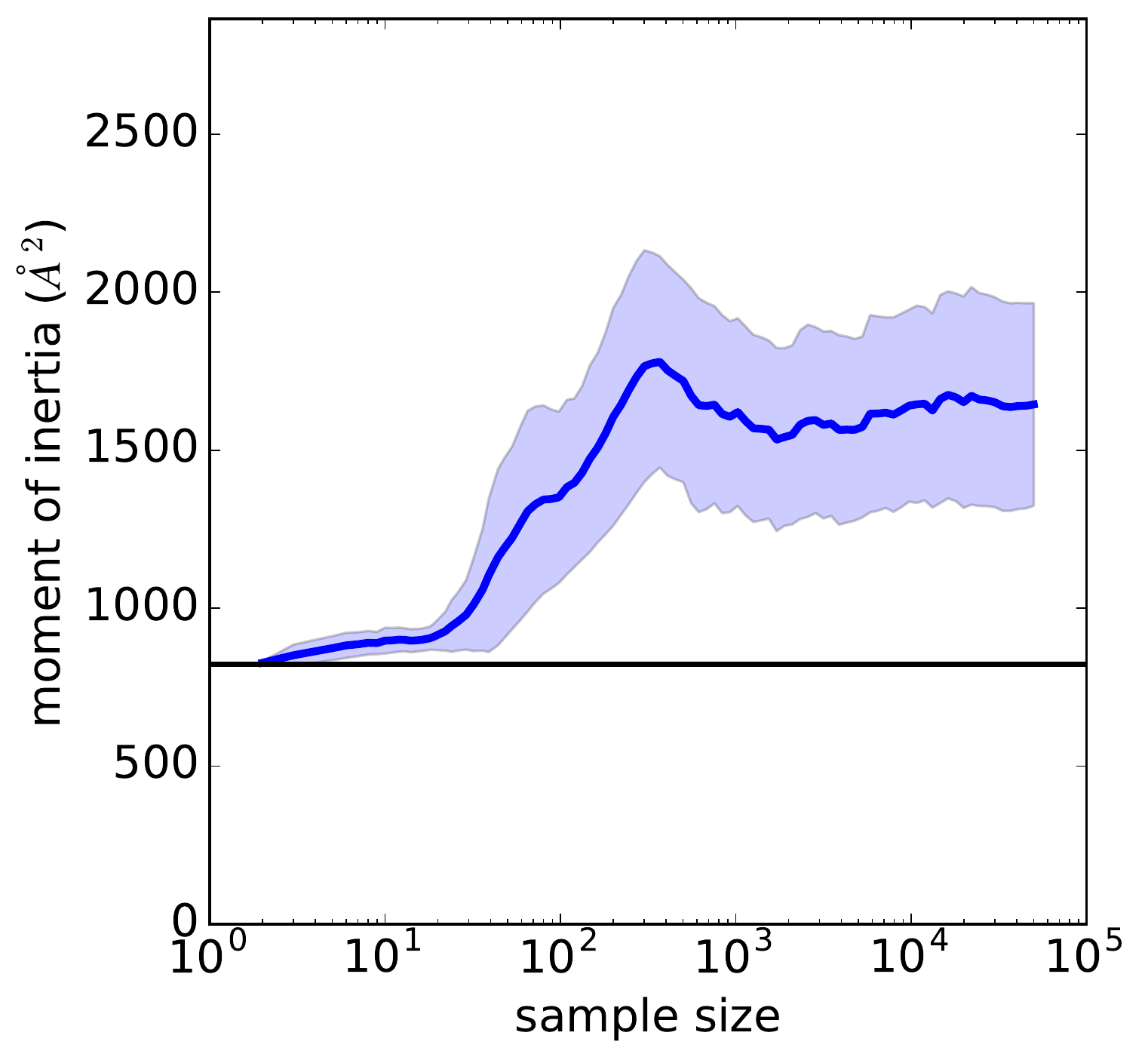}
\caption{\label{convergence3}Distribution of moment of inertia throughout our GP method as the sample size is increased. The solid blue line traces the mean, and the shaded area is one standard deviation of the distribution either side. Compare with the histogram in Figure 3a of the main text. After about 1000 samples, the distribution does not change significantly.}
\end{figure}


\begin{figure}[b]
\includegraphics[width=8cm]{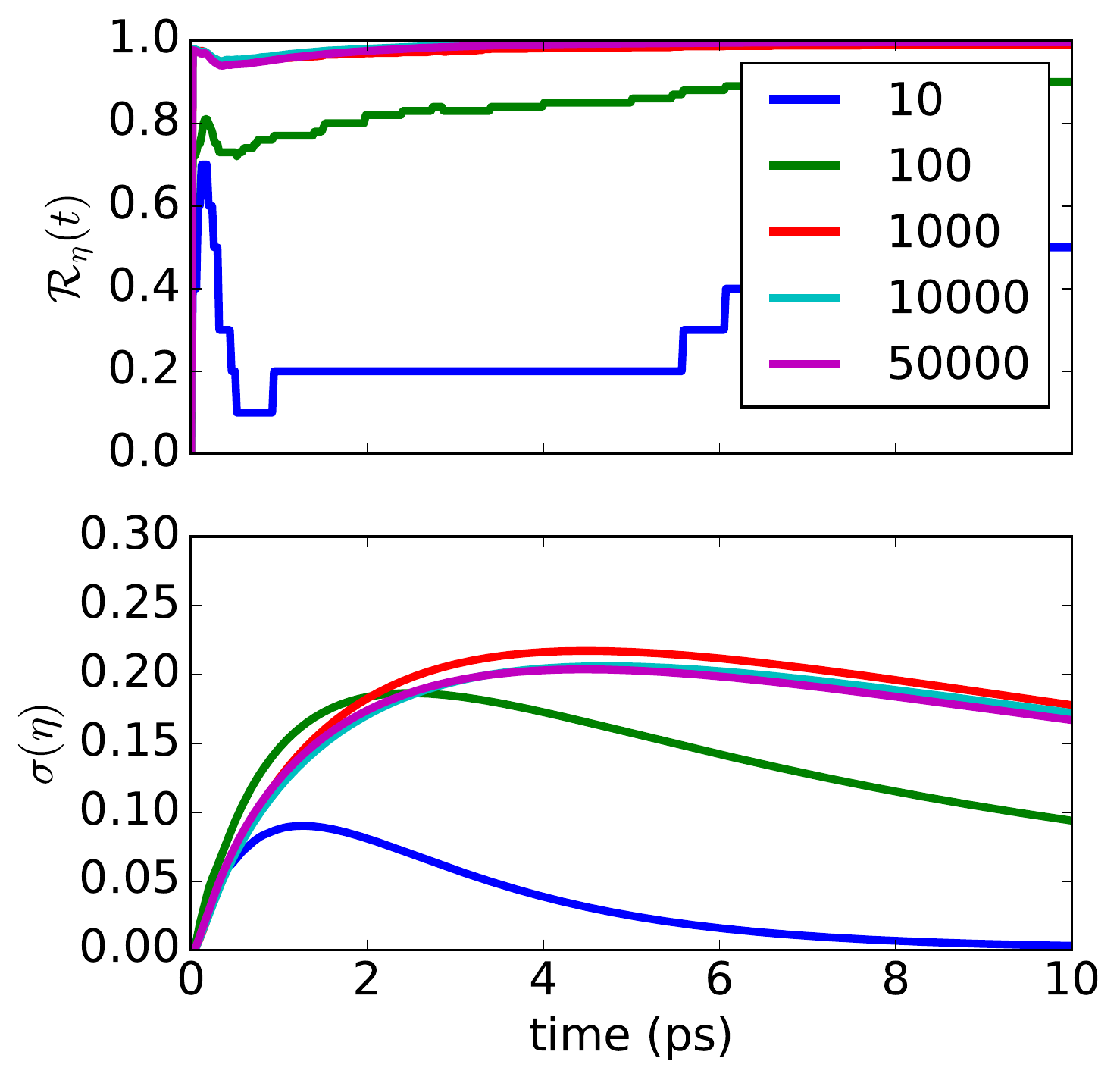}
\caption{\label{convergence1}Time-resolved rank and standard deviation of our GP method as the sample size is increased. }
\end{figure}
\begin{figure}[t]
\includegraphics[width=8cm]{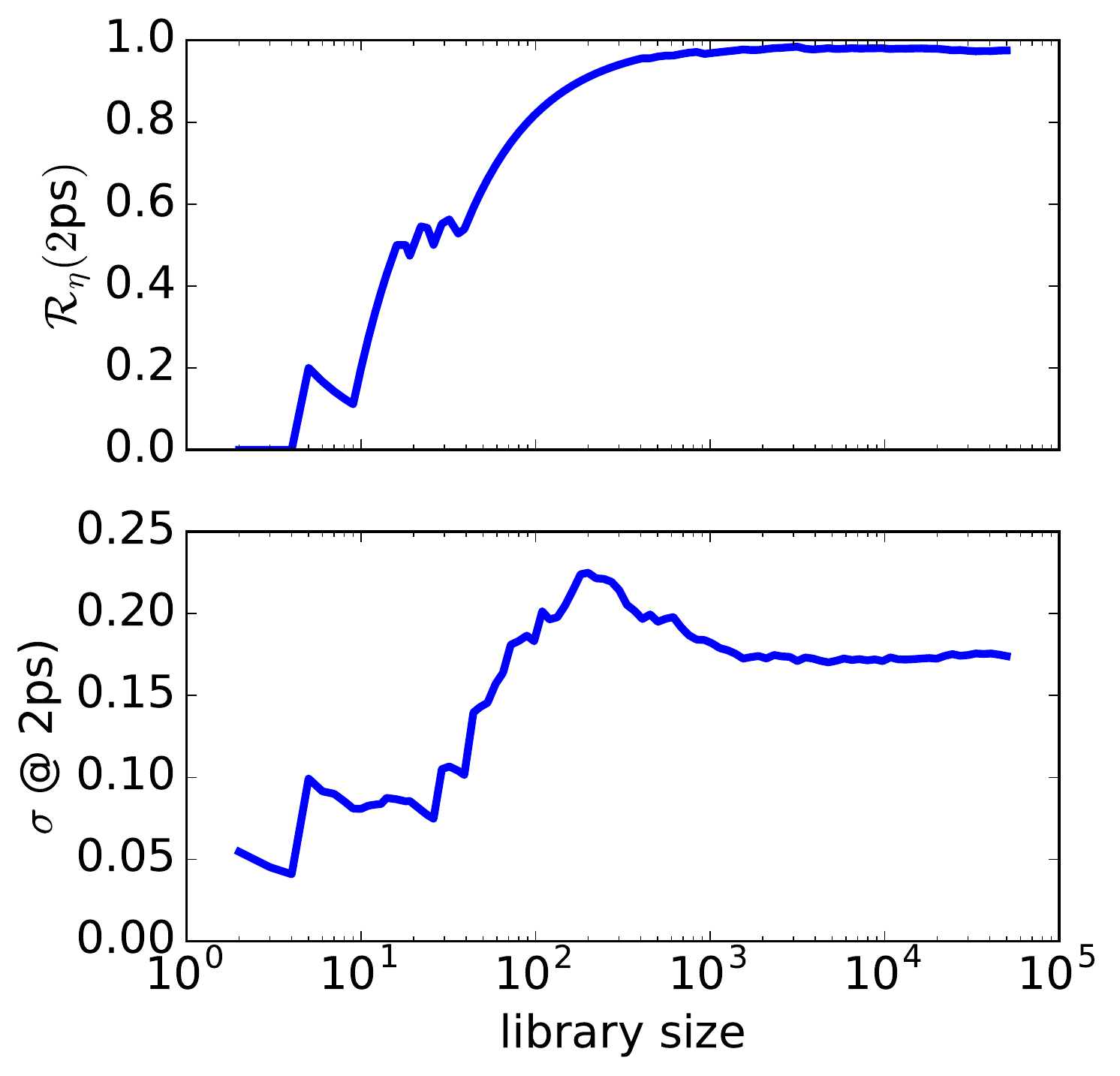}
\caption{\label{convergence2}Convergence diagnostics for rank and standard deviation (evaluated at 2ps) of the GP method. Note how the values `settle' after the sample size is increased above 1000.}
\end{figure}

\section{Spearman correlation}
Our quantifier of correlation is chosen as the Spearman rank correlation coefficient~\cite{Spearman1904,Zwillinger1995}
\begin{align}
\mathcal{S}(x,y;t) &= \frac{R_{xy}(t)}{\sqrt{R_{xx}(t)R_{yy}(t)}};\\
R_{xy}(t) &= \sum_i \mathcal{R}_x(i,t)\mathcal{R}_y(i,t)\nonumber\\
&\phantom{=}-\left(\sum_i \mathcal{R}_x(i,t)\right)\left(\sum_i \mathcal{R}_y(i,t)\right)\nonumber
\end{align}
It takes values in $[-1,1]$, and measures the extent to which the two variables $x$ and $y$ are monotonically related (unlike the more common Pearson correlation coefficient, which measures the extent to which $x$ and $y$ are linearly related). The Spearman quantity is usually defined with the ordinal rank, but it is simple to prove that using the normalised rank (as in the expression above) is equivalent. The Spearman coefficient is exactly the Pearson coefficient of the rank of $x$ with the rank of $y$. We also made use of the idea of a p-value to judge the \emph{significance} of the Spearman quantity (not to be confused with the strength of the correlation). The p-value is the probability of seeing a correlation at least as strong as the one we saw, but in the case when the two variables are completely uncorrelated. In our studies this quantity was always less than $1\times10^{-45}$. 
\section{More detail on structure and ATC method.}
 The x-ray crystal structure of the natural FMO complex, \emph{Prosthecochloris aestuarii} variety was determined to $1.3\mbox{\normalfont\AA}$ resolution in Ref.~\cite{TronrudWenGay2009}. Our reference structure, containing only the chromophores, was obtained by superimposing the DFT optimised structure of the chromophores in vacuum on the crystal structure from the Protein Data Bank~\cite{3EOJ,BermanWestbrookFeng2000}. This structure was the starting point for the GP procedure.

The Atomic Transition Charges (ATCs) have been computed on the density-functional theory (DFT)- optimised structure of the isolated chromophore in vacuum. Excited state calculations used time dependent DFT, the long range corrected LC-$\omega$PBE exchange-correlation hybrid functional and a 6-31G** Gaussian type basis set. For the LC-$\omega$PBE functional, an $\omega$ cutoff of $0.3A^{-1}$ was used, as this has been found to minimise the root-mean-square deviation of the energies of a number of ground and excited states for numerous medium sized molecules~\cite{VydrovScuseria2006,RohrdanzMartinsHerbert2009,FranclPietroHehre1982}. The matrix elements {$V_{mn}^{(i)}$ are scaled by the factor {$f=0.8$ to capture the dielectric effect of protein and solvent~\cite{AdolphsMuhMadjet2007}.

Writing the reference Hamiltonian in the chromophore basis, the output of our ATC calculation yields:

\begin{align}
H_1=\left(
\begin{array}{rrrrrrrr}
 11549.72 &  -154.04 &     8.91 &    -8.89 &    10.18 &   -12.39 &   -11.38 &    27.72 \\
  -154.04 & 11412.76 &    43.29 &     9.94 &     3.05 &    10.41 &     5.63 &     6.72 \\
     8.91 &    43.29 & 11331.78 &   -88.06 &    -0.02 &   -12.50 &     8.97 &     0.97 \\
    -8.89 &     9.94 &   -88.06 & 11436.72 &  -120.75 &   -26.18 &   -80.94 &    -2.46 \\
    10.18 &     3.05 &    -0.02 &  -120.75 & 11436.72 &    61.71 &    -6.35 &     5.27 \\
   -12.39 &    10.41 &   -12.50 &   -26.18 &    61.71 & 11517.78 &    46.32 &   -14.56 \\
   -11.38 &     5.63 &     8.97 &   -80.94 &    -6.35 &    46.32 & 11500.76 &   -15.86 \\
    27.72 &     6.72 &     0.97 &    -2.46 &     5.27 &   -14.56 &   -15.86 & 11485.76 \\
\end{array}
\right)\text{(cm)}^{-1}.
\end{align}
The largest magnitude of coupling is about $154$ (cm)$^{-1}$, which (using $\hbar=5.3$ps/cm) is equivalent to about $39$ps$^{-1}$, meaning that (in the absence of other dynamics) this matrix element would be responsible for coherently exchanging a quantum of energy between chromophore 1 and chromophore 2 every $25$fs . As stated in the main text, the isolated-chromophore energies (diagonal matrix elements) are taken from Ref.~\cite{JiaMeiZhang2015} (obtained there via a QM/MM calculation). We attribute the discrepancy between the off-diagonal elements of this matrix and others in widespread use in the literature (most frequently reproduced from Ref.~\cite{AdolphsRenger2006}) to a significant difference in approach. Our calculation proceeds from more recent x-ray crystallography data~\cite{TronrudWenGay2009}, and is characterised by the atomic transition charge method.  
Note that we consider the ATCs $q^{t}_{\alpha m}$ to be a fixed property of the chromophore, which does not change between structures. This is justified since the relative positions of the chromophores would result in only a small perturbation to the excited state charge density. This allows us to avoid expensive excited state electronic structure calculations, facilitating the collection of a much larger set of ersatz structures.
\section{Unbiased $M_i$ sample}
By altering the acceptance criterion in our GP algorithm, we were able to construct a sample of ersatz structures that have a distribution of $M_i$ (moment of inertia) that is tightly constrained (variance less than 1  $\mbox{\normalfont\AA}^2$) and (importantly) unbiased with respect to the reference value $M_1$ (to a very good approximation). We label this sample GP(C), see Figure \ref{MOIdist}. The results show a decreased rank $\mathcal{R}$ for the reference structure, supporting the view that $M_i$ is an important structural feature (see Figure \ref{tight}). The rank is still high however, showing that the reference benefits from some other structural motifs, too. These include the $M_i^{(\ddag)}$,$DO_i$ and $DO_i^{(\ddag)}$ quantities discussed in the main text.

\begin{figure*}[t!]
\includegraphics[width=16cm]{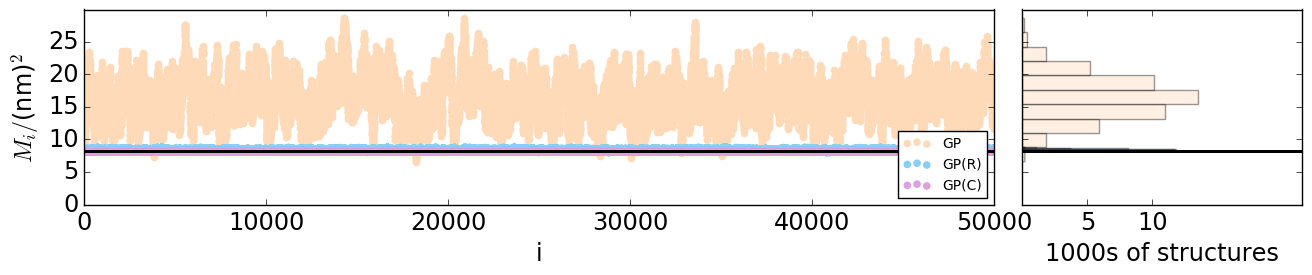}
\caption{\label{MOIdist}Moment of inertia distributions for GP, GP(R) and GP(C) samples. }
\end{figure*}

\begin{figure}[t!]
\includegraphics[width=6cm]{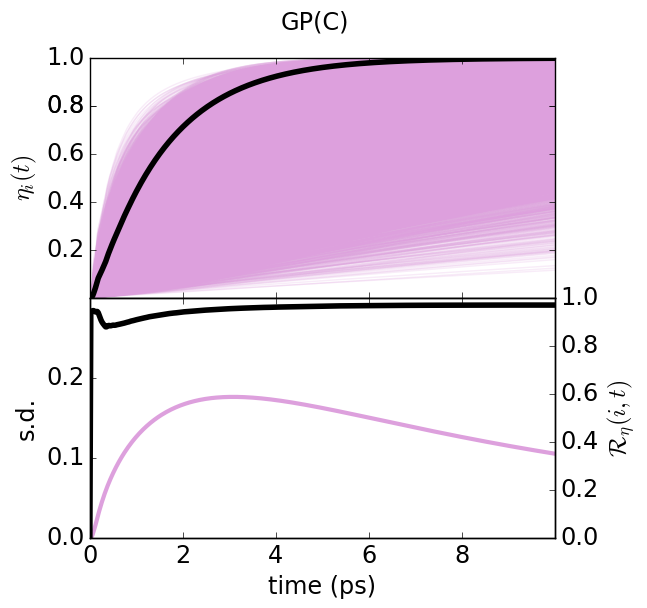}
\caption{\label{tight}Compared to our original GP sample, the reference structure ranks slightly lower in a sample where the distribution of moment of inertia $M_i$ has been engineered to be very narrow and unbiased around the reference value $M_1$. We label this sample GP(C) for `constrained'.}
\end{figure}

\begin{figure}[t!]
\includegraphics[width=6cm]{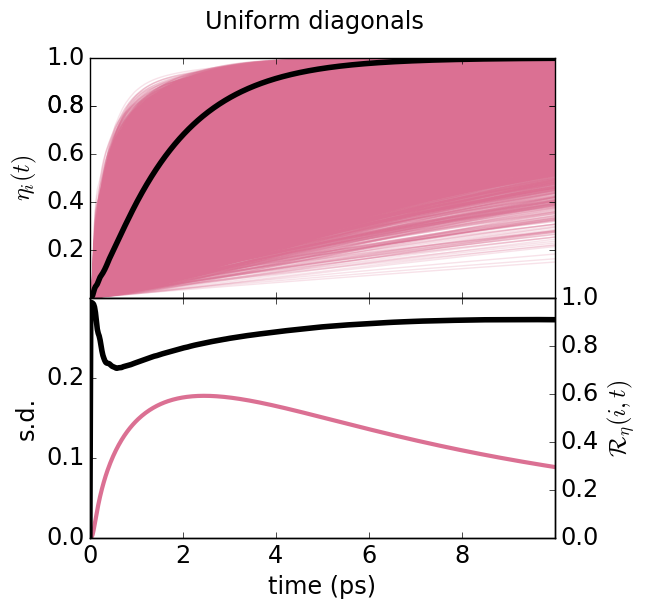}
\caption{\label{uniform}Compared to our original GP sample, the reference structure ranks slightly lower in a sample where the energy distribution $\epsilon_l$ is set to uniform.}
\end{figure}

\section{Uniform energy sample}
To explore the relative importance of the reference FMO's energy distribution $\epsilon_l$, we cloned our GP sample but then forced all sites to have an energy of $11549.72$ (cm)$^{-1}$. The results are shown in Figure~(\ref{uniform}). We also noted by visual inspection that 9 of the 10 best structures (at 5 ps) had one or more chromophores clearly between source and drain. 

\begin{figure}[t!]
\includegraphics[width=8cm]{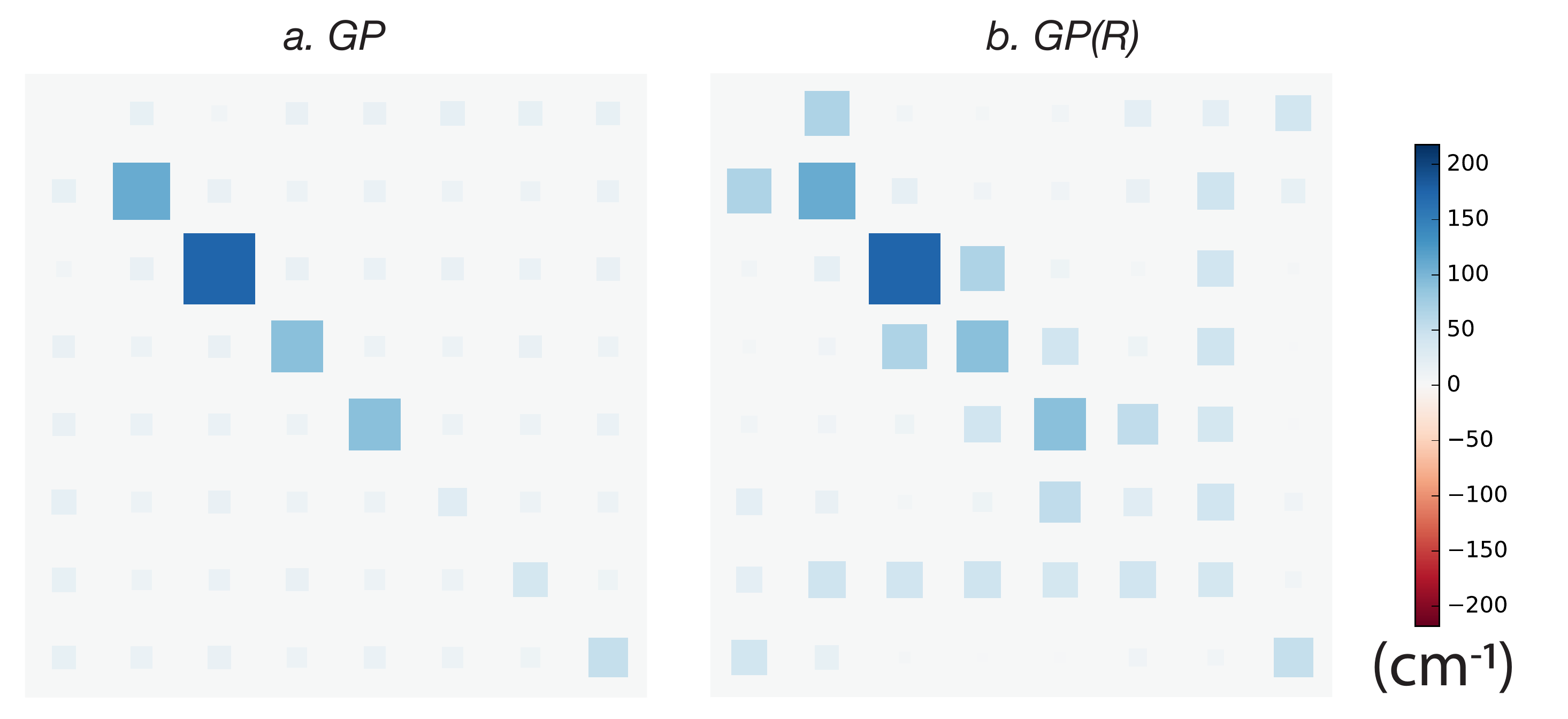}
\caption{\label{averagecoupling} The average absolute value of the 64 matrix elements of the excitonic Hamiltonian under a) full geometric perturbation (GP) and b) rotations only (GP(R)).}
\end{figure}

\section{Average coupling strengths}
Figure~(\ref{averagecoupling}) shows the average $\langle |H_i|\rangle$ for GP and GP(R) approaches. The notation $|H_i|$ implies that the absolute value of each matrix element is taken. The later (rotations only) method clearly benefits from larger overall couplings on average.
 
\section{Uncorrelated matrix element perturbation (UMEP)}
UMEP is a technique commonly employed to generate alternative Hamiltonian matrices by independently changing matrix elements~\cite{CarusoChinDatta2009,BakerHabershon2015}. Particularly with the off-diagonal elements, this is already potentially a mistake, since it cannot be mirroring physical changes such as those that we consider: those must be correlated when viewed in the chromophore basis. It is not possible to change the coupling between a first chromophore and a second, without also changing the coupling with all other chromophores. UMEP methods are also usually implemented with the matrix-element changes normally distributed around zero, with a small standard deviation. The matrix-element changes behave like errors in the model parameters and propagate through to a distribution in the exciton dynamics. It is therefore not surprising that the final distribution is also centred on the original dynamics $\mathcal{R}(t)\approx0.5$, and has a small standard deviation $\sigma < 0.06$ ($\sim 4 \times$ less than our GP method). See Figure \ref{UMEP}, where we implement the UMEP approach from~\cite{BakerHabershon2015}: off-diagonal Hamiltonian matrix elements get increased by a random number, distributed normally with mean zero and standard deviation 3.5 cm$^{-1}$. Diagonal elements get increased by a random number, distributed normally with mean zero and standard deviation $[25.5,42.5,25.5,25.5,51,51,51,51]\textrm{cm}^{-1}.$ The reasoning is as follows: suppose that the site energies that apply to the reference geometry itself represent a reasonable estimate of the site energies which may be present in any `FMO-like' protein. 

As we motivate in the main paper, our GP method is a much better motivated way to alter the couplings of a reference structure. A similarly motivated `physically plausible' method of changing the on-site energies is highly desirable, but outside the scope of this study. In the interim, we make use of a restricted form of UMEP to investigate the importance of energetics versus couplings. The restricted UMEP only changes diagonal matrix elements, which correspond to the energy of well separated chromophores.  Whilst we find it more acceptable to have these changing in an uncorrelated way, we proceed only cautiously since the problems of sampling changes physically, with respect to bias (is it `easier' to make arrange for higher on-site energies than lower ones?) and variance  (is it `easier' to make a shallow energy gradient than a steep one?) remain. 

Using restricted UMEP, we study the complementary situation to the main text: there, we looked at fixing the energies and changing the couplings, finding the FMO couplings to be well attuned. It is interesting to ask: is it also true that the energies of the natural FMO are well attuned when the couplings are fixed? We find that they are not, since in such a library $\mathcal{R}(t)\approx0.5$. See Figure \ref{UMEPdiag}.

\begin{figure}[t!]
\includegraphics[width=6cm]{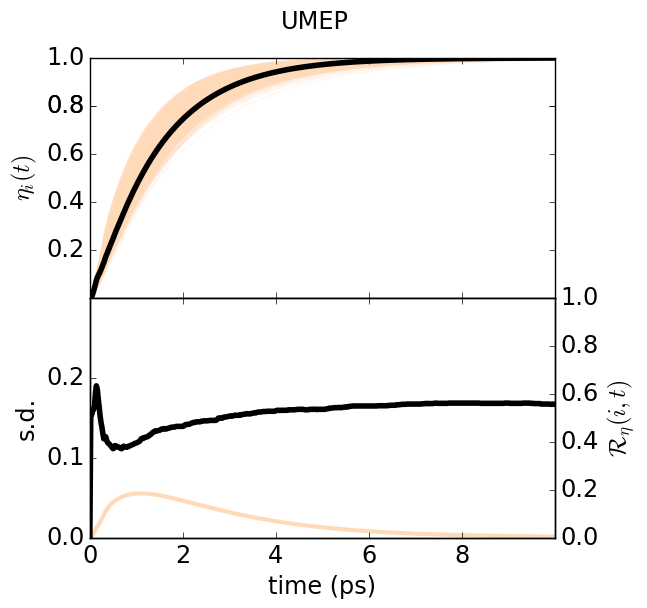}
\caption{\label{UMEP}Uncorrelated Matrix-Element Perturbation methods tend to give a lower standard deviation of transport properties, and have the natural FMO ranking in the middle of the class. }
\end{figure}

\begin{figure}[t!]
\includegraphics[width=6cm]{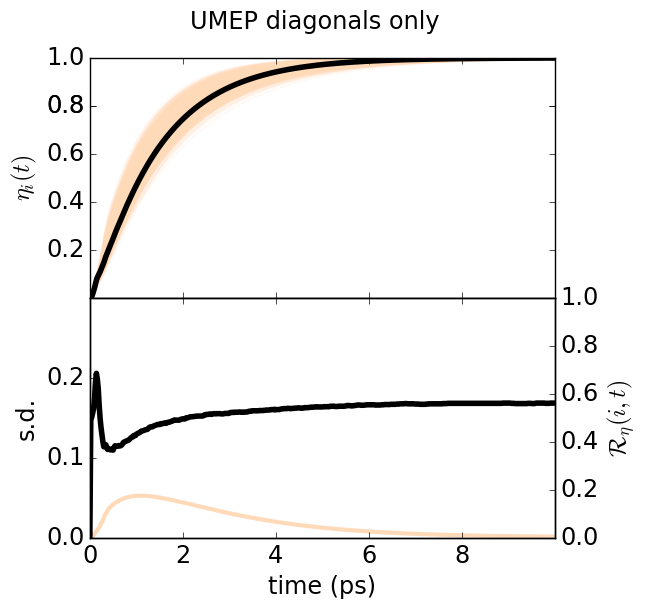}
\caption{\label{UMEPdiag}Randomly varying the diagonal elements of the reference Hamiltonian leads to a roughly equal number of higher and lower performing matrices, and once more the natural FMO is ranking in the middle of the class. Hence with a fixed geometry, there is scope for speeding up exciton transport by altering the isolated chromophore energies. Contrast with Figure 3 of the main text, where conditional on the natural FMO energies, the structure was found to be very high ranking.}
\end{figure}

\end{document}